\documentclass[prl,twocolumn,preprintnumbers,amsmath,amssymb,superscriptaddress]{revtex4}

\usepackage{graphicx}
\usepackage{dcolumn}
\usepackage{bm}
\usepackage{color}
\usepackage{pifont}
\newcommand{\cmark}{\ding{51}}%
\newcommand{\xmark}{\ding{55}}%

\begin{document}


\title{Exciton-polariton condensates}
\author{Tim Byrnes}
\affiliation{National Institute of Informatics, 2-1-2 Hitotsubashi, Chiyoda-ku, Tokyo 101-8430, Japan}

\author{Na Young Kim}
\affiliation{E. L. Ginzton Laboratory, Stanford University, Stanford, CA 94305}

\author{Yoshihisa Yamamoto}
\affiliation{National Institute of Informatics, 2-1-2 Hitotsubashi, Chiyoda-ku, Tokyo 101-8430, Japan}
\affiliation{E. L. Ginzton Laboratory, Stanford University, Stanford, CA 94305}

\date{\today}

\begin{abstract}
Recently a new type of system exhibiting spontaneous coherence has emerged -- the exciton-polariton condensate. Exciton-polaritons (or polaritons for short) are bosonic quasiparticles that exist inside semiconductor microcavities, consisting of a superposition of an exciton and a cavity photon.  Above a threshold density the polaritons macroscopically occupy the same quantum state, forming a condensate. The lifetime of the polaritons are typically comparable to or shorter than thermalization times, making them possess an inherently non-equilibrium nature.   Nevertheless, they display many of the features that would be expected of equilibrium Bose-Einstein condensates (BECs). The non-equilibrium nature of the system raises fundamental questions of what it means for a system to be a BEC, and introduces new physics beyond that seen in other macroscopically coherent systems.  In this review we focus upon several physical phenomena exhibited by exciton-polariton condensates.  In particular we examine topics such as the difference between a polariton BEC, a polariton laser, and a photon laser, as well as physical phenomena such as superfluidity, vortex formation, BKT (Berezinskii-Kosterlitz-Thouless) and BCS (Bardeen-Cooper-Schrieffer) physics.  We also discuss the physics and applications of engineered polariton structures.
\end{abstract}

\maketitle


Spontaneous coherence is a phenomenon that has fascinated physicists from a wide range of fields, ranging from condensed matter physics, atomic physics, quantum optics, to high-energy physics.  Lasing is
the most familiar phenomenon giving rise to macroscopic coherence, in this case formed by stimulated emission of photons \cite{sargent78}. Bose-Einstein condensation (BEC) is another example of collective coherence of many particles, such that above a critical density (or equivalently below a critical temperature), the particles spontaneously occupy the ground state \cite{pitaevskii03}.
The superfluidity of liquid $^4\mbox{He}$ is a manifestation of spontaneous coherence in the presence of strong interactions \cite{tilley90}. Superconductivity, viewed as a condensation of Cooper pairs, allows for a charged version of BEC yielding resistanceless (superfluid) flow \cite{leggett06}.   In this review, we examine a new system that undergoes spontaneous coherence: exciton-polariton condensates.   The recent observation of exciton-polariton condensation \cite{kasprzak06,balili07,deng02} adds another particle to the list for which BEC has been observed  -- cold atoms \cite{anderson95,davis95}, magnons \cite{nikuni00,demokritov06}, and more recently photons \cite{klaers10}. We shall see that rather than being simply another type of particle that undergoes BEC, it possesses characteristics that incorporates new physics due to
their intrinsically non-equilibrium nature.

One of the distinctive features of exciton-polaritons (or simply polaritons for short) is its exceedingly light effective mass, typically of the order of $10^{-4} $ times the bare electron mass. For an ideal (non-interacting and at thermal equilibrium) bosonic gas in three dimensions, the critical temperature for BEC occurs when $ n \lambda_D^3 = 2.62 $, where $ n $ is the density of the bosons and $ \lambda_D = \sqrt{\frac{2 \pi \hbar^2}{m k_B T }} $ is the de Broglie wavelength ($ m $ is the mass of the bosons, $ k_B $ is Boltzmann's constant, and $ T $ is the temperature).  This criterion can be intuitively be thought of as when the density of the bosons is high enough such that their wavefunctions overlap.  The mass dependence of  $\lambda_D$ means that it is easier to produce BECs for light mass particles, which is one of the great advantages of using polaritons.  For most experiments to date, exciton-polariton condensates are produced at cryogenic temperatures in the vicinity of $ \sim 10  $ K using materials such as GaAs and CdTe. However, using other materials such as GaN, ZnO, and organic semiconductors, polariton condensates at higher temperatures including room temperature have been also realized \cite{christopoulos07,baumberg08,kenacohen10,guillet11}. This makes exciton-polaritons a fascinating topic not only from a fundamental perspective but also of potential practical interest to future quantum technological devices. As will be discussed in more detail, exciton-polariton condensates form in effectively two-dimensional structures rather than three dimensions. This allows for investigating interesting physics peculiar to two dimensions such as the BKT (Berezinskii-Kosterlitz-Thouless) transition where there is an interplay of long-range order and thermally excited vortices.

In this review, we give an overview of the recent rapid progress made in the realization and understanding of exciton-polariton condensates.  As there have been several reviews \cite{deng10,kavokin10,richard10,snoke10,keeling11,timofeev12,carusotto13} focusing primarily on the fundamental aspects of exciton-polaritons, here we shall emphasize some of the more recent developments relating to exciton-polariton condensates. In particular we will discuss physical phenomena exhibited such as superfluidity, vortex formation, nontrivial phases of polariton condensates exhibiting BKT and BCS (Bardeen-Cooper-Schrieffer) physics, and
the physics and applications of engineered polariton structures.  One important issue which arises invariably when discussing exciton-polaritons is the role played by the open-dissipative nature of the system. From the early investigations of coherence formation in polaritons it has been a controversial issue of whether the concept of an exciton-polariton BEC is valid at all from a thermodynamical perspective.  We will discuss the known differences between a standard photon laser, a polariton laser, and a polariton BEC, and discuss how the non-equilibrium character of the system changes the nature of the physical effects observed.

\section{Exciton-polaritons: Basic aspects}

We first describe briefly some basic aspects of exciton-polaritons.  For a more detailed exposition we refer the reader to reviews such as Refs. \cite{deng10,kavokin10,richard10,snoke10,keeling11,timofeev12,carusotto13}.  We shall primarily give examples for the most widely used GaAs and CdTe-based systems for the sake of concreteness,  although other materials are conceptually similar but with different parameter values.
 A typical planar microcavity for exciton-polaritons consists of several quantum wells (QWs) sandwiched by two distributed Bragg reflectors (DBR) shown in Fig. \ref{fig1}a. A QW is a thin layer (typically of the order of 10 nm) of a relatively narrow bandgap material (such as GaAs or CdTe) surrounded by a wider bandgap material (doped with Al or Mg respectively).
The QW exciton is the primary excitation of the system under external stimulation, for example by illumination from a laser, as a Coulombically bound electron-hole pair.
The excitons are strongly confined in the growth ($z$-) direction due to the QW but are free to move in the $x$-$y$ plane. Excitons, by virtue of being  composite particles made of two fermions, obey bosonic statistics as long as their density is low enough such that they do not overlap (the typical Bohr radius of a QW exciton is $ \sim 10 $ nm) \cite{yamamoto99}. The use of several QWs allows to distribute the total exciton density over the QWs to avoid reaching this saturation (or Mott) density, while simultaneously enhancing the cavity coupling.  Excitons have typical lifetimes of the order of
 100 ps to 1 ns in GaAs, CdTe semiconductors.

While excitons already have a relatively light effective mass by atomic standards ($ \sim 0.2 m_e$, where $ m_e $ is the bare electron mass), by the use of a cavity it is possible to make them considerably lighter \cite{weisbuch92}. Photons are confined within the cavity in the $z $-direction, while they are free to move in the $x$-$y$ plane (Fig. \ref{fig1}a). This gives them an effective mass typically of the order $ \sim 10^{-4} m_e $.   The effect of the cavity is to create strong coupling between the mobile excitons and photons, resulting in new quasi-particles which are a superposition of the two: exciton-polaritons.
The resulting dispersion relations are shown in Fig. \ref{fig1}b. The photon and exciton dispersions anticross under strong coupling, resulting in two new dispersion relations for the lower polariton (LP, lower energy branch) and upper polariton (UP, higher energy branch).   The Rabi splitting $ \Omega $ of exciton-polaritons  (defined as the energy difference between the LP and UP dispersion at zero in-plane momentum $ k $) is typically of the order of $ \sim 10 $ meV (in multi-QW GaAs and CdTe samples), and is larger or comparable to the relevant system parameters such as the temperature, binding energy, and interactions.  This allows us to consider to good approximation the quasiparticles of the system to be exciton-polaritons, rather than the original excitons or photons.  The exciton-polariton effective mass and the lifetime are predominantly determined by their photonic component, whose effective mass is much lighter and lifetime is much shorter than the excitonic parts. The photons have a finite lifetime due to leakage of the light through the microcavity mirrors. Samples that were used in early experiments had lifetimes typically of the order of $ \tau \sim $ 1 ps, however more recently these have been extended to
$ \tau \sim$ 10-100 ps \cite{nelsen13,tanese13}. While the polariton lifetimes are of the order of the photon, the polariton-polariton interactions are inherited from their excitonic component, arising primarily from Coulomb exchange effects of the underlying fermionic species
\cite{schmittrink85,ciuti98}.  This gives a typical mean field interaction of the order of $ \sim $ meV, resulting in extremely light quasiparticles with short lifetimes, but simultaneously possessing a sizable nonlinear interaction.

Condensation then occurs by the following process. First a population of polaritons must be introduced, which requires excitation from an external source, usually a laser (Fig. \ref{fig1}b).  Since the aim is to show  spontaneous coherence of the polaritons, they should be introduced in such a way as the original coherence of the laser is lost. This can be done primarily by two methods: (1)  resonant pumping to exciton energy at a large angle (Fig. \ref{fig1}b) and (2) non-resonant pumping at the reflection minimum of the stop band at $ k=0 $.  In either method, a hot cloud of polaritons is produced mostly occupying the LP dispersion.  The LPs then undergo scattering with the crystal and dissipate their energy via phonon emission. This cooling happens efficiently down to momenta such that the exciton and photon energy difference is of the order of the Rabi splitting, which is where the photonic component of the polariton becomes appreciable \cite{tassone99}. At this momentum the lifetime of the polariton is greatly reduced, such that less time is available for cooling. Simultaneously the dispersion steepens due to the Rabi splitting at this momentum, reducing the phonon density of states making the cooling less efficient.  This creates a ``bottleneck'' in the polariton population, and a second mechanism becomes responsible for further cooling of the polaritons.  For a sufficiently dense population of the
bottleneck polaritons, polariton-polariton scattering takes over as the dominant cooling process. For example, two polaritons within the bottleneck region can scatter, leaving one in the vicinity of $ k=0 $ and another at twice the bottleneck momentum.  The higher energy polariton can cool again via phonon emission, resulting in a lower overall energy of the cloud. This secondary cooling mechanism allows for momentum states within this bottleneck region to build up a macroscopic population. The population buildup occurs most likely at the $ k=0 $ state by virtue of its lowest energy. We note that other pumping schemes in addition to the above two methods have been used in several works, most notably the optical parametric oscillator (OPO) scheme \cite{spano12}.

Some typical measurement data of condensation, by Kasprzak, Devaud, and co-workers, are shown in Fig. \ref{fig1}c \cite{kasprzak06}.
Here we see the energy-momentum dispersion relation, directly measured from the photoluminescence emerging from the DBR mirrors.  Below condensation, there is a broad distribution of exciton-polaritons in both energy and momentum of the LP dispersion.  As the pumping power is increased, there is a sudden narrowing in both the energy and momentum distribution and a large population of polaritons occupies the zero momentum mode of the system, consistent with the formation of a BEC. Figure \ref{fig1}d plots the typical dependence of the polariton population with the pump power.  Below threshold, the population at $ k = 0 $ increases linearly with the pump power, but increases nonlinearly with the onset of condensation \cite{deng03}.  The population dependence returns to a linear dependence after the transition region, since now the condensate population dominates the total polariton population.  The macroscopic occupancy of the ground state, the nonlinear threshold behavior, and the narrowing of the linewidth are all properties that would be expected of BEC.  However, these alone are not sufficient to conclude that BEC has occurred, and one requires careful cross checking with other expected properties.  This is the topic of the next section.

\section{Is it a laser or a BEC?}

One of the most controversial issues relating to exciton-polariton BECs, is whether it should be called a BEC at all.  Looking again at the device structure in Fig. \ref{fig1}a, 
it consists of a cavity enclosing a QW  which supports electron-hole excitations - precisely the same structure as a vertical-cavity surface emitting laser (VCSEL). In the VCSEL the electrons and holes act as the gain medium and lasing occurs via population inversion where many electron-hole pairs are excited.  Coherent light is emitted by the usual process of stimulated emission where the light in the cavity is amplified by recombination of electron-hole pairs.  In the procedure described in the previous section of creating the polariton condensate, we excite large population of high energy excitons to supply the condensate.  Could not the coherent light emitted by the polariton BEC be more simply interpreted as standard lasing, where the excitons play the role of the gain medium? Another question relates to the short lifetime of the polaritons, which means that the condensate must be continually replenished in order to have a stable population.  Since the usual concept of a BEC assumes thermodynamic equilibrium, in such an explicitly non-equilibrium situation, does it make sense to even think of a polariton BEC? While some of these issues are subject of ongoing debate \cite{butov07,butov12,devaud12}, we describe the present state of understanding addressing these questions.

The first difference between a polariton BEC and a VCSEL is in which particle species accumulates coherence \cite{imamoglu96,snoke12} (Fig. \ref{fig2}a).  From a device perspective, one clear difference between a polariton BEC and a VCSEL microcavity structure is the presence or absence of strong coupling respectively.  In the lasing case, the gain medium (electron-hole pairs) is pumped sufficiently such that population inversion occurs. Then via a process of stimulated emission, photons are emitted and amplified coherently such that eventually lasing occurs. Thus the species that develops coherence is the photons and the gain medium is not coherent.  In contrast, in a polariton BEC a large population of hot uncondensed polaritons is initially excited.  Assuming that the polariton lifetime is sufficiently long for thermalization with the crystal to occur, by a process of stimulated scattering into the $ k=0 $ mode in a similar way to Fig. \ref{fig1}b, a polariton condensate forms.   The polaritons that are in the condensate then emit coherent light via leakage
of their photonic components through the microcavity mirrors.  In this case, the coherence that accumulates is in the polaritons rather than the cavity photons.  We note that the respective photonic and excitonic fractions do not matter \cite{kasprzak08}, and we can consistently talk about either extreme --  photon BECs \cite{klaers10} and exciton BECs.  Thus even though both the laser and the polariton condensate emit coherent light, in this case there is a clear distinction determined by what particle species becomes coherent.

While the above distinction is clear if polariton lifetimes are very long relative to the thermalization time, this is not always the case in practice, making the distinction less obvious. In the literature it has now become commonplace to refer to different regimes in various ways.  On one end of the spectrum is the {\it polariton BEC}, which is the ideal case discussed above, where the lifetimes are long enough such that thermalization occurs.  At the other end of the spectrum strong coupling is lost, which may arise in a variety of different ways such as short lifetimes or additional dephasing. This is referred to as a {\it photon laser}. An intermediate regime, where strong coupling and macroscopic occupation of the polariton ground state is present, but without a thermalized population of polaritons, is referred to a {\it polariton laser}.  The polariton laser was originally introduced as a novel type of laser where population inversion was not necessary to form a coherent polariton population \cite{imamoglu96,dang98}.  The term {\it polariton condensate} is used as a broader term to encompass both the polariton BEC and polariton laser.  In the early experiments the polariton lifetimes were on the order of  $ \tau \sim$  1 $\mbox{ps} $ and the stimulated scattering times of a similar order or less \cite{kasprzak06,deng06}. Nevertheless, thermalization of the polaritons following a Boltzmann exponential decay of the distribution function was observed at threshold (Fig. \ref{fig2}b). At higher excitation powers, there is an increase in population of particles at lower energy deviating from the exponential distribution, as would be expected of a BEC. However, fitting with a Bose-Einstein distribution is typically difficult, as the chemical potential and the temperature of the polaritons are unknown parameters, and non-equilibrium effects may cause deviations from the Bose-Einstein distribution. 

Thermalization is only one of many characteristics that is expected in a BEC. Table \ref{tab1} gives a checklist of characteristics for a polariton BEC, a polariton laser, and a photon laser.
For a BEC off-diagonal long-range order is a central concept, where there is the appearance of a macroscopic wavefunction $ \psi(r) $ forming an order parameter \cite{pitaevskii03}.  Experimentally this corresponds to extended spatial coherence $ g^{(1)} (r) $ across the condensate above threshold, but short ranged correlations below threshold -- a fact confirmed in experiments such as Refs. \cite{kasprzak06,deng07}.  In addition to the spatial
coherence, the polaritons should occupy one of the spin degrees of freedom macroscopically (optically active polaritons have spin $ \pm 1$), which is observed as a linear polarization of the emitted light above the condensation threshold \cite{deng03,kasprzak06}. For cases where there is no bias in any spin direction, the polarization emerges stochastically, and the spin symmetry is spontaneously broken \cite{baumberg08}. For a macroscopic wavefunction, the position and momentum uncertainty product is of the order of the Heisenberg limit, as confirmed in Ref. \cite{deng07}. Other effects that are
expected are an increase in temporal coherence, originating from the spectral narrowing of the condensate, and confirmation that condensed particles are polaritons as opposed to cavity photons.
As discussed above, the detection of whether the coherent particle species is a polariton or a photon is at the heart of distinguishing between various types of coherence. However, spectroscopic experimental data does not always provide unambiguous evidence as several contributing factors may affect the energy of the polariton. In such cases
the measurement of a Zeeman shift by an externally applied magnetic field has been used to identify the existence of an excitonic component to the polaritons, distinguishing it from a photon laser \cite{schneider13}.

Despite the progress in the realization and understanding of a polariton BEC, questions remain. The buildup of spontaneous coherence between the two extremes of the polariton BEC and photon laser \cite{assmann11}
suggests that it may be more appropriate to think in a unified way of the continuum between a laser, a distinctly non-equilibrium phenomenon, and an equilibrium BEC.  To investigate this aspect there has been increased activity in the identification of a second threshold above the polariton BEC transition. At pumping powers considerably above the condensation threshold (typically about 10-100 times), observations appearing to be photon lasing have been known to occur for some time \cite{dang98,tempel12}.  Theoretically the mechanism of this is now only beginning to be understood \cite{yamaguchi13}.
Another question relates to the steady leakage and replenishment of the polaritons to maintain a constant condensate population.  
This potentially changes the properties of the BEC, such as superfluidity and its various quantum phases in a fundamental way. How do we understand the physics of BECs in this new non-equilibrium setting?  Some of these aspects will be discussed in the following sections, but are an ongoing topic of investigation.

\section{Superfluidity}

Superfluidity is the phenomenon of flow without friction, and is often observed in systems that undergo spontaneous coherence, such as BECs \cite{pitaevskii03}, liquid $^4\mbox{He}$ \cite{tilley90}, and superconductors \cite{leggett06}.  It allows for spectacular effects such as persistent currents that flow indefinitely, the impossibility of rotating the superfluid, and flow through capillaries without viscosity. Central to the concept of superfluidity is the Landau criterion which gives the maximum velocity $ v_c = \min_k \varepsilon_k /k $ that the fluid can flow while still maintaining its superfluid properties, where $ \varepsilon_k  $ is the energy-momentum dispersion.  While the Landau criterion is typically derived from general principles of Galilean transformations \cite{pitaevskii03}, it is also possible to understand its origins directly from a simple scattering argument \cite{carusotto06}. For a moving condensate with a parabolic dispersion, particles within the condensate always have other states with the same energy to scatter into. The scattering eventually destroys the single momentum state occupation the condensate is originally in, and eventually acts to slow the average motion.  However, if the moving condensate has a linear dispersion arising from Bogoliubov interactions, and if the velocity is less than the Landau velocity, then there are no states that are resonant to scatter into.  This suppression of scattering then acts to maintain the condensate velocity, even in the presence of scattering defects, resulting in superfluidity.

How well does this argument transfer to polariton condensates?  Again, the non-equilibrium nature makes the straightforward application of this argument problematic. Experimentally, evidence exists showing polariton condensates possess a linear Bogoliubov dispersion 
\cite{utsunomiya08,kohnle11}.  However in the more general case it is widely believed that the excitation spectrum is modified due to the open-dissipative nature of condensate.  The model, introduced by Wouters and Carusotto, which is often used to describe the system, is a dissipative Gross-Pitaevskii (GP) equation with gain and loss terms, coupled to an incoherent reservoir  \cite{wouters07}.  The reservoir is a dynamical variable, which can modify the dispersion of the condensate rather strongly.  The energies of the dissipative GP eigenstates also become complex in general, reflecting the decaying nature of the excitations.
A typical excitation spectrum for a homogeneous condensate potential is shown in Fig. \ref{fig3}a. Instead of the lowest-lying excitations starting linearly
around $ k = 0 $, they instead begin only at a non-zero value \cite{byrnes12}. Thus the main characteristic of the Bogoliubov dispersion, the linearity around zero momentum, is not necessarily present in polariton BECs (although for particular parameters it can exist).  
The problem is that when the Landau criterion is applied to such a dispersion, it always gives $ v_c =0 $, implying that superfluidity never occurs.  However,
this naive application of the Landau criterion contradicts numerical evidence showing suppressed Rayleigh scattering \cite{carusotto04}, persistent current flow and suppressed drag \cite{wouters10}.  Figure \ref{fig3}b shows a numerical calculation of condensate flow at various velocities in the presence of a defect. We see that despite the dissipative nature of the GP equation, there is a characteristic velocity that the fringes due to the defect disappear, equal to the sound velocity (the gradient of the Bogoliubov dispersion at $ k = 0 $) $ c_s = \sqrt{g |\psi|^2 / m } $ where $ g $ is the interaction energy, $ |\psi|^2 $ is the density and $ m $ is the polariton mass. This occurs even despite the lack of any definable Landau velocity in the dispersion.  One of the reasons for this discrepancy may be attributed to the pump/decay dynamics of the polaritons.  Unlike standard superfluidity where scattering of polaritons leads to an eventual slowing down of the condensate, in a polariton condensate the scattered polaritons may quickly decay off, with the pump reinforcing the dominant condensate mode.

From an experimental point of view, several works have shown evidence consistent with superfluidity in exciton-polariton condensates.  The most compelling evidence of superfluidity to date is the measurement of Rayleigh scattering of a polariton condensate in the presence of an impurity \cite{amo09b,amo09}.  In the experiments performed by Amo and co-workers, a polariton condensate is excited by a continuous wave laser, and centered around natural defects that exist in the sample.  As shown in Fig. \ref{fig3}c, as the density is increased the amount of Rayleigh scattering due to the condensate flow diminishes. The effect of increasing the density is to increase the critical velocity according to the Landau criterion, as the sound velocity $ c_s $ increases with density.  At low density, the distribution of the polaritons in momentum space has strong scattering to momenta of equal magnitude, a clear signal of Rayleigh scattering, while at higher densities the scattering is suppressed as would be expected in a superfluid. In the experiments of Sanvitto, Bloch, and co-workers \cite{sanvitto10}, first a polariton condensate is prepared, then a 2 ps pulsed laser in a Laguerre-Gauss mode carrying angular momentum is imprinted on the condensate to induce a vortex state.  The circular flow induced by the vortex was seen to survive for long times
 limited only by the uncontrollable random walk within the condensate which the vortex core undergoes. 

The current state of experiments and theory suggest that there are effects consistent with superfluidity, in agreement with the simple Landau criterion \cite{keeling09}. However, it appears further work is required to completely understand this and the notion of superfluidity in this non-equilibrium setting. Another related question regarding excitation spectrum is the observation of the ``ghost'' branch of the Bogoliubov spectrum which has been predicted to exist theoretically \cite{wouters09,byrnes12}. In addition to the standard positive Bogoliubov spectrum, a negative branch should be visible in the photoluminescence. This was observed in the OPO regime \cite{kohnle11}, but to date there has been no direct photoluminescence measurement.

\section{Phases of polariton condensates}

In an infinite two dimensional (2D) system, off-diagonal long-range order associated with BEC in non-interacting quantum degenerate Bose gas breaks down since it is vulnerable to thermal fluctuations at non-zero temperatures \cite{hohenberg67,mermin66}. However, a 2D interacting Bose gas in an infinite system is predicted to exhibit quasi-long range order preserving superfluidity by spontaneously forming vortex pairs via the BKT transition \cite{berezinskii72,kosterlitz73}. Above the BKT critical temperature, the quantum gas excites free single vortices induced by phase fluctuations. However, when the system cools down, the vortices pair up and phase fluctuations in space cancel out, reducing the free energy, and recovering quasi long-range coherence and superfluidity.
BKT physics has been one of the central themes in the investigation of low-dimensional quantum fluids in recent years \cite{hadzibabic11}. In particular, there is still incomplete understanding of the quantum phase of an interacting Bose gas in a finite sized 2D system. Exciton-polariton superfluids are a good testing ground to examine this unresolved problem, and active theoretical and experimental efforts have examined the gain-loss mechanism and the dynamics of vortices via time-resolved and phase-resolved imaging techniques.

A quantized vortex is a topological defect, with zero density at its core and a multiple of $ 2 \pi $ phase rotation around it. Single vortices and vortex pairs in exciton-polariton condensates excited by non-resonant pumping have been observed either pinned at defects or imprinted from the phase fluctuation of lasers.  Vortices are detected via interferometry, where the spatial phase and intensity distributions are reconstructed.  The first observation of vortices in exciton-polaritons was reported in CdTe systems \cite{lagoudakis08}.  In this experiment, Lagoudakis and co-workers identified a fork-like interference pattern, which is the signature of a phase dislocation at an accidental local potential minimum (Fig. \ref{fig4}a).  The same team also observed pinned half quantum vortices using polarization-resolved interferograms, which shows a $\pi $-rotation in phase and a $\pi $-rotation in polarization \cite{lagoudakis09}.  Figure \ref{fig4}b images a single vortex-antivortex pair observed in a GaAs system, induced by an inhomogeneous spot profile of the pump laser \cite{roumpos11}. Whereas all the above works are under an incoherent pumping scheme, vortex-antivortex (V-AV) pairs have also been observed in the OPO regime \cite{tosi11}. Here, a pump-probe technique allows one to track the evolution of the V-AV pairs. The experimental observations have been consistent with numerical calculations using a dissipative GP equation. While the observation of vortices and V-AV pairs may suggest a connection to a BKT phase, in fact the current experiments are not yet conclusive of this. In a true BKT phase, vortices and V-AV pairs associated with phase fluctuations are thermally activated.  One way to identify the BKT phase would be via a single shot measurement, but the low signal-to-noise ratio of short-lived polaritons is a present limitation.

Another way to explore the BKT phase is to characterize the spatial correlation function $g^{(1)}(r)$, which quantifies the off-diagonal long-range order of the system. The BKT phase is predicted to show a power-law decay of $g^{(1)}(r)$, exhibiting quasi-long range spatial coherence.  This is possible if bound V-AV pairs are created where the phase disturbance of a vortex is exactly canceled out with that of an antivortex. Thermodynamically, V-AV pairs are favorable for decreasing the free energy by increasing the entropy. While quantitative measurements of $g^{(1)}(r)$ in quantum fluids are missing, exciton-polariton systems are promising to establish this directly by interferometric measurements. Several reports of $g^{(1)}(r)$ in one-dimensional exciton-polaritons \cite{wertz10,manni11} indicate that long-range order exists above threshold. According to
equilibrium BKT theory the long-range power law decay should have a characteristic exponent of 0.25. The first measurements
indicated a higher value in the range 0.9-1.2 \cite{roumpos12}, but later measurements gave better agreement once technical imperfections such as pump noise and fragmentation were eliminated \cite{nitsche14}. Smoking-gun experiments to elucidate the BKT-BEC crossover are the single-shot direct observation of free vortices above the BKT temperature, bound V-AV pairs in the BKT phase, and a vortex-free regime
at lower temperatures in the BEC phase.  While short-lived polaritons ($\tau \sim 1$ ps) are unlikely to show this evidence conclusively, the recent advent of long-lived polaritons ($\tau \sim 10$-$100$ ps) \cite{nelsen13,tanese13} are good candidates for showing this crossover and superfluidity.

Solitons are another type of topological excitation present in polariton BECs. Repulsive polariton-polariton interactions are predicted to support dark solitons \cite{pigeon11}, which is characterized by a phase slip in the propagation of exciton-polariton fluids. Such solitary waves may be produced by disturbing the condensate with a localized obstacle.  Via an interferogram,  the phase along the soliton trajectory was mapped as shown in Fig. \ref{fig4}c \cite{amo11}.  Since the first report of the observation of exciton-polariton solitons, the temporal dynamics of dark solitons \cite{grosso12} and bright solitons at the lower-polariton inflection with negative mass \cite{sich12} have been observed. More recently, it was shown that polaritons in the linear regime (i.e. no polariton-polariton interactions) could also reproduce 
the intensity and phase patterns that are expected of solitons purely from interference effects \cite{cilibrizzi14}.  The question of what is 
necessary for the unambiguous claim of observing a soliton is a topic of current debate.

Another type of crossover that has been considered for polaritons condensates is the BEC-BCS crossover.  For exciton BECs, it has been long theoretically predicted that such a crossover should exist as the density is varied \cite{keldysh68,comte82}. The physical picture of this crossover is that at low densities the electrons and holes within the exciton are relatively strongly bound by their mutual Coulomb attraction, and condense at sufficiently low temperatures due to their bosonic nature.  At higher densities, their mutual attraction becomes screened by the large population of electrons and holes, and form loosely correlated Cooper pairs, described by a BCS wavefunction.  While this picture is well established for exciton BECs, does the same hold of polariton BECs? This question was first examined by Keeling, Littlewood, and co-workers \cite{keeling04,keeling05} and followed up by several others using a BCS wavefunction approach \cite{byrnes10,kamide10}.  The consistent picture which emerges from these studies is that at high densities the photons completely dominate the dynamics in what can be described as a photon BEC state \cite{klaers10}. The reason for this is simple: the electrons and holes have a maximum allowable population due to the Pauli exclusion principle since they are fermions, while photons do not have this restriction.  Therefore at some point the number of photons overwhelm the excitons so that the photons dominate the properties of the system.  
The photons have the effect of producing an effective attractive interaction induced by the Rabi coupling between the electrons and holes.  
Therefore, increasing the density in polariton condensates acts generally in the opposite direction to the exciton BEC-BCS crossover:  the attraction between the electrons and holes in the excitons is reinforced by the Rabi coupling.  
These studies show that the presence of the photons considerably alter the quantum state of the system at sufficiently high densities, even without the effect of the cavity decay. Taking this in account, we may expect that with a sufficiently short photon lifetime all these regimes are more suitably thought of as a laser, giving a crossover between photon BEC, polariton BEC, electron-hole BCS, and photon lasing regimes.
The full understanding of this phase diagram is an active topic of current research \cite{kamide10,yamaguchi13}.

\section{Engineered Polariton Structures}

The realization of the optical lattice in cold atomic gases has allowed for the unprecedented ability to manipulate BECs, with seminal experiments such as the superfluid-Mott insulator transition \cite{greiner02},
kicking off the field of quantum simulations \cite{buluta09}.  Likewise, the manipulation of the quantum state of the exciton-polariton condensate is a important task that is necessary for both investigating fundamental physics such as the BKT transition and realizing future polaritonic quantum devices.
Figure \ref{fig5}a shows a zoo of engineering methods to create static and dynamical in-plane potentials by modifying either the photonic or excitonic modes, which consequently affects the polariton states.  The techniques include chemical etching for forming pillars and strips \cite{wertz10,tanese13}, laser pump spot formation \cite{tosi12}, piezoelectric acoustic lattices \cite{mendez10}, and thin metal film deposition \cite{lai07,kim11}.  Single trap configurations have realized  energy separation
of the $s$-orbital ground state from higher-orbital states \cite{kaitouni06,nardin10}.  Other geometries that have been realized include coupled molecules forming bonding and antibonding states \cite{galbiati12},
and one- \cite{lai07,tanese13} and two-dimensional lattices \cite{kim11,masumoto12,kim13,kusudo13} exhibiting band structures. For one-dimensional lattices,
antiphased $p$-orbital states were realized at the Brillioun zone edges \cite{lai07}, a similar effect to that known in weak links of superfluid $^3$He forming metastable $p$-states \cite{backhaus98}.  Due to the negative effective mass and the repulsive interparticle interaction at the Brillioun zone edge, a
bright soliton was observed inside the gap, as shown in Fig. \ref{fig5}b \cite{tanese13}.  The excited state condensation in lattices  results from the interplay of gain-relaxation dynamics and the finite lifetime of the polaritons, making it more favorable to condense in these states.  The same physics applies to 2D lattices, where the metastable $d_{xy}$-orbital condensation appears at high symmetry $M$ points in a square lattice \cite{kim11}. Other geometries such as the honeycomb lattice \cite{kusudo13}  have been realized, where V-AV lattices have been observed, as wells as a linear dispersion
due to Dirac points in triangular lattices \cite{kim13}. Figure \ref{fig5}c shows the Brillouin zones of a honeycomb lattice potential above the polariton condensation threshold. Laser pump spot engineering can control the condensate wavefunction to produce 2D lattices, where a stable V-AV honeycomb lattice has been observed (Fig. \ref{fig5}d) \cite{tosi12}.

The creation of periodic potentials suggest applications to quantum simulation, where complex quantum many-body systems are realized that are beyond the reach of computer simulations \cite{buluta09}.  Another possibility in this direction is the creation of metamaterials that have no natural realization, with exotic Hamiltonians. Hybridizing different orbital symmetry of coherent exciton-polaritons would be a basis to search for bosonic orbital order. Furthermore, by incorporating spin dynamics controllable by light polarization, exciton-polaritons have the potential to explore magnetic order which could  provide insights to the behavior of fermions in strongly correlated systems. Other types of applications include polaritronic circuits (the polariton analogue of atomtronic circuits) \cite{deveaud08,amo10,ngyuen13,ballarini13} and novel light sources \cite{imamoglu96,byrnes13}. The recent achievement of electrical pumping, replacing the laser pumping methods described here, give the potential for ultra-low threshold lasers \cite{schneider13,bhattacharya13,imamoglu96}.

\section{Outlook}

We have seen that exciton-polariton condensates exhibit a rich variety of phenomena, possessing characteristics in common to atomic BECs,
but in a new hitherto unexplored non-equilibrium regime. Even the most fundamental question of whether it is a BEC or a laser forces us to reanalyze what we mean exactly when we make these distinctions.
As we suggested at the beginning of this review, the phenomenon of spontaneous coherence appears to be a rather robust concept, stretching between the superficially distinct quantum states of superconductors,
lasers and BECs.  Perhaps what exciton-polariton condensates teach us is that a generalization is required to unify these concepts together, rather than categorization into each of regimes.
We have seen that the cavity photon loss and the quasi-bosonic nature of the exciton modify familiar physical effects such as superconductivity, vortex formation, BKT and BCS phases, giving the opportunity to re-examine these from a new perspective \cite{snoke08}.

As discussed throughout this review there are many open problems that are still ongoing topics of investigation. We have seen that the mechanism of non-equilibrium superfluidity, two-threshold crossover to photon lasing, evidence of BKT and BCS physics, and signatures of solitons are some of the growing list of questions being further investigated currently.  Perhaps one of the most fascinating prospects this system has comes from the ability of engineering the microcavity configuration by nanofabrication. This suggests applications in quantum simulation, where quantum lattice models can be investigated in an engineerable system.
Many of the fabrication methods allow in principle for an arbitrary device geometry, not limited to periodic or simple trap configurations, and technological improvements in the quality of the samples will undoubtedly continue to occur.  This control in device structure combined with room temperature operation makes polariton condensates attractive for future quantum technological applications.  While currently it is still unclear what the major application of polariton condensates will be, the fundamental questions that it opens will undoubtedly give us a better understanding of the phenomenon of spontaneous coherence, and the remarkable physics which comes with it.

\bibliographystyle{naturemag}

\section*{Acknowledgments}
We thank B. Devaud-Pl{\'e}dran for providing valuable comments on the manuscript. This work is supported by the FIRST program through JSPS, the Okawa Foundation, the Transdisciplinary Research Integration Center, and DARPA QuEST program through Navy/SPAWAR Grant N66001-09-1-2024.

\section*{Competing financial interests}
The authors declare no competing financial interests.

\clearpage

\begin{widetext}


\begin{center}
\begin{table}[t]
\begin{tabular}{cccc}
\hline
Property & Exciton-polariton  & Exciton-polariton  & VCSEL \\
 & BEC & laser  &  \\
\hline
Thermal equilibrium below threshold \cite{kasprzak06,balili07} & \cmark  & \xmark  & \xmark \\
Bose distribution above threshold & \cmark & \xmark  & \xmark  \\
Threshold corresponds to onset of degeneracy \cite{kasprzak06,balili07,deng02} & \cmark & \cmark & \xmark  \\
Linewidth narrowing \cite{kasprzak06,balili07} & \cmark & \cmark  & \cmark \\
Increase of temporal coherence $g^{(1)} (\tau) $ \cite{kasprzak06,love08}  & \cmark & \cmark  & \cmark \\
Spontaneous polarization \cite{baumberg08} & \cmark &  ?  & \xmark \\
Long range spatial coherence $g^{(1)} (r) $ \cite{kasprzak06,balili07,deng07}   & \cmark & \cmark & \cmark \\
Polaritons are the particles that accumulate coherence (strong coupling) \cite{schneider13} & \cmark & \cmark & \xmark \\
Heisenberg limited position and momentum uncertainty product \cite{deng07} & \cmark & \cmark & \xmark \\
\hline
\end{tabular}
\caption{\label{tab1}  Differences between an exciton-polariton BEC, exciton-polariton laser, and a VCSEL (Vertical Cavity Surface Emitting Laser). The properties that have been experimentally demonstrated for polariton condensates are shown as the references.}
\label{table1}
\end{table}
\end{center}

\begin{figure}
\includegraphics[width=\textwidth]{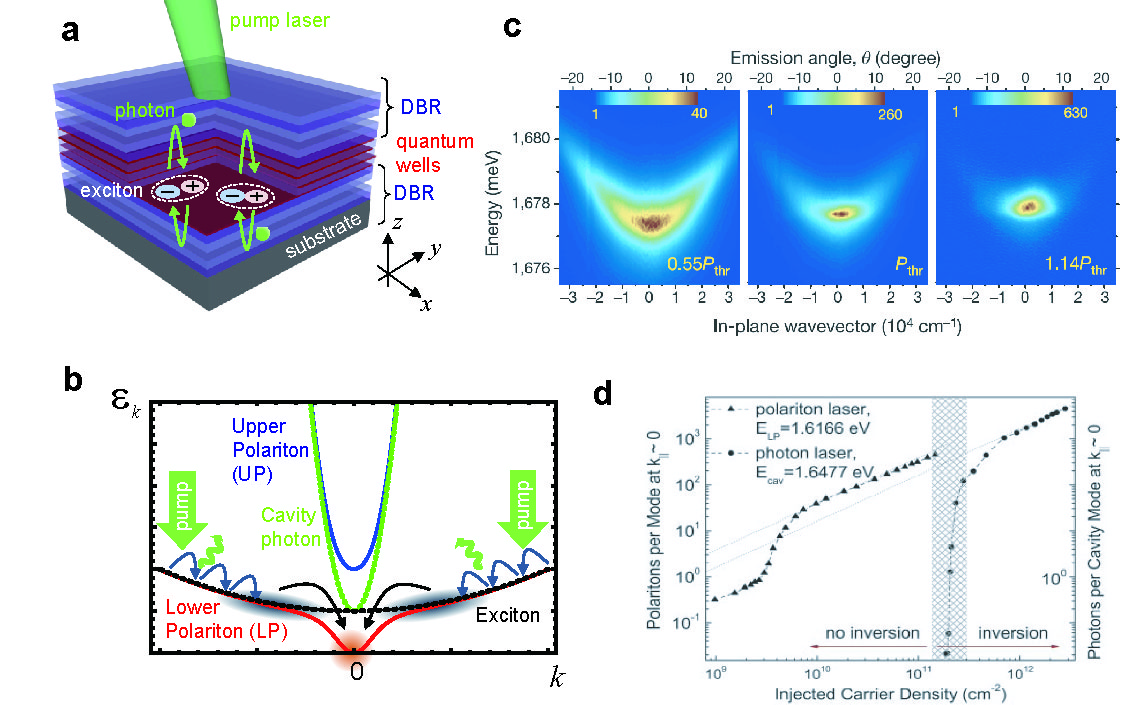}
\caption{Exciton-polariton condensation.  {\bf a} Typical device structure supporting exciton-polaritons. Excitons, consisting of a bound
electron-hole pair, exist within the quantum well layers.  These are sandwiched by two Distributed Bragg Reflectors (DBRs), made of
alternating layers of semiconductors with different refractive indices.  The DBRs form a cavity which strongly couples a photon and an exciton
to form an exciton-polariton. Polaritons are excited by a pump laser incident from above.
{\bf b} Exciton-polariton dispersion and condensation process.  Strong coupling between the cavity photon and exciton dispersions split the dispersions near $ k = 0 $ to create the lower polariton (LP) and upper polariton (UP) dispersions.  The pump laser initially excites high energy excitons which cool via phonon emission towards the bottleneck region (black cloud).  Excitons in the bottleneck region then scatter into the
condensate via stimulated cooling.  {\bf c} Experimental dispersion images of polariton condensate formation from Ref. \cite{kasprzak06}.  Below the threshold for condensation the polaritons are broadly distributed in momentum and energy.  At and above threshold the polaritons condense in the $k = 0 $ ground state. {\bf d} Polariton ground state population for a polariton laser as a function of the pump power from Ref. \cite{deng03}.  Figure also shows the threshold for a standard laser achieved by a sufficiently large detuning to lose strong coupling in the same sample for comparison.
\label{fig1}}
\end{figure}

\begin{figure}[t]
\includegraphics[width=\textwidth]{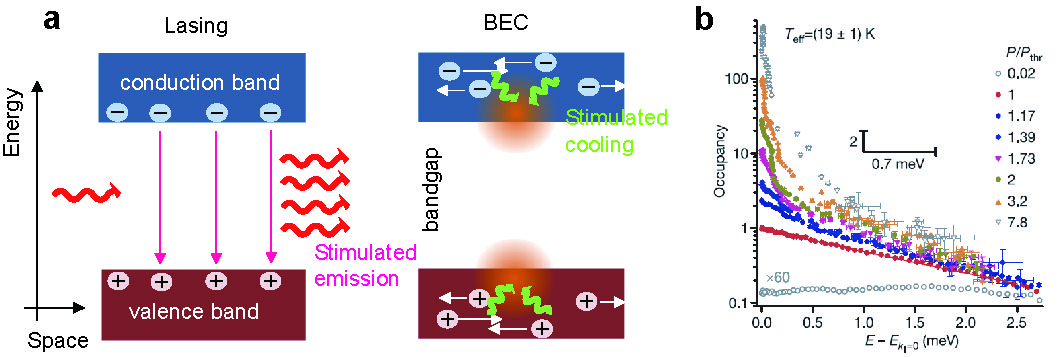}
\caption{Lasing versus condensation.  {\bf a} Distinction between photon lasing and a polariton BEC. In lasing coherence is formed via stimulated relaxation of electron hole pairs into a cavity photon mode.  An polariton BEC forms via stimulated cooling of hot polaritons into a low energy states.  {\bf b} Experimental measurement of the polariton occupancy distribution in Ref. \cite{kasprzak06} at and above the $ k =0 $ ground state for various pump powers.  At threshold the distribution has a exponential Boltzmann-like distribution indicating thermalization.
\label{fig2}}
\end{figure}

\begin{figure}[t]
\includegraphics[width=\textwidth]{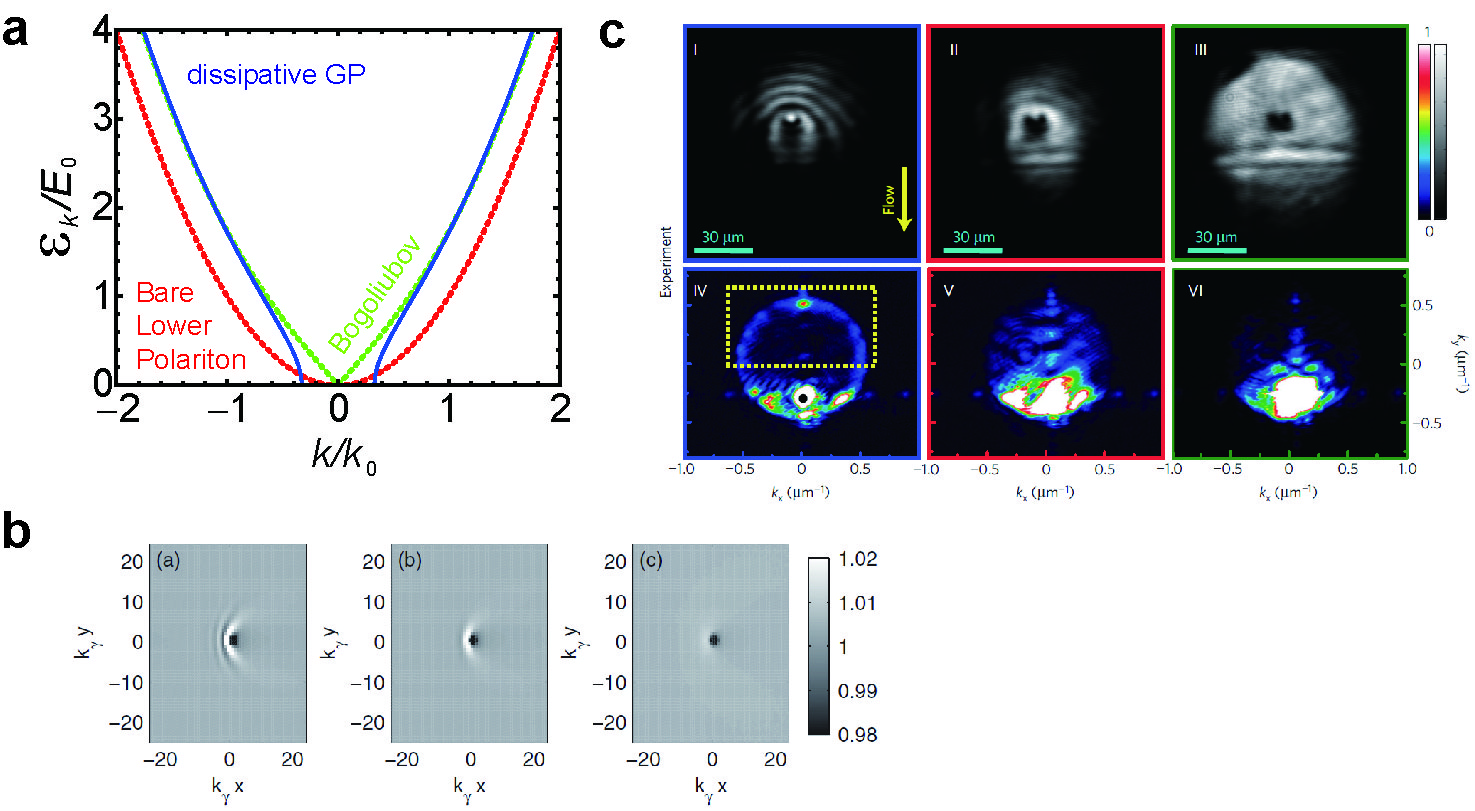}
\caption{Superfluidity in exciton-polariton condensates. {\bf a} Excitation spectrum of the dissipative Gross-Pitaevskii (GP) equation. The bare lower polariton dispersion (no interactions) and the Bogoliubov dispersion is shown with the dashed lines for comparison. Momentum is normalized to the experimental length scale $ k_0 $ and energies $ E_0 = \frac{\hbar^2 k_0^2}{2m} $. Typical numbers are for example $ k_0 = 1 {\mu m}^{-1} $ and $ E_0 = 0.68 \mbox{meV} $.  Parameters used are the same as Fig. 1(a) in Ref. \cite{byrnes12}.  {\bf b} Numerical simulation of the density a moving condensate in real space with a single defect using the dissipative GP equation in Ref. \cite{wouters10}. The velocity of the condensate is (a)  $ v/c_s = 1.5 $ (b) $ v/c_s = 1$ (c)  $ v/c_s = 0.4$.  {\bf c} Experimental real (top row) and momentum (bottom row) space images of a polariton condensate impacting on a defect.  As the velocity of the condensate is reduced (left to right), in real space the fringes disappear, while in momentum space Rayleigh scattering become suppressed. \label{fig3} }
\end{figure}

\begin{figure}[t]
\includegraphics[width=\textwidth]{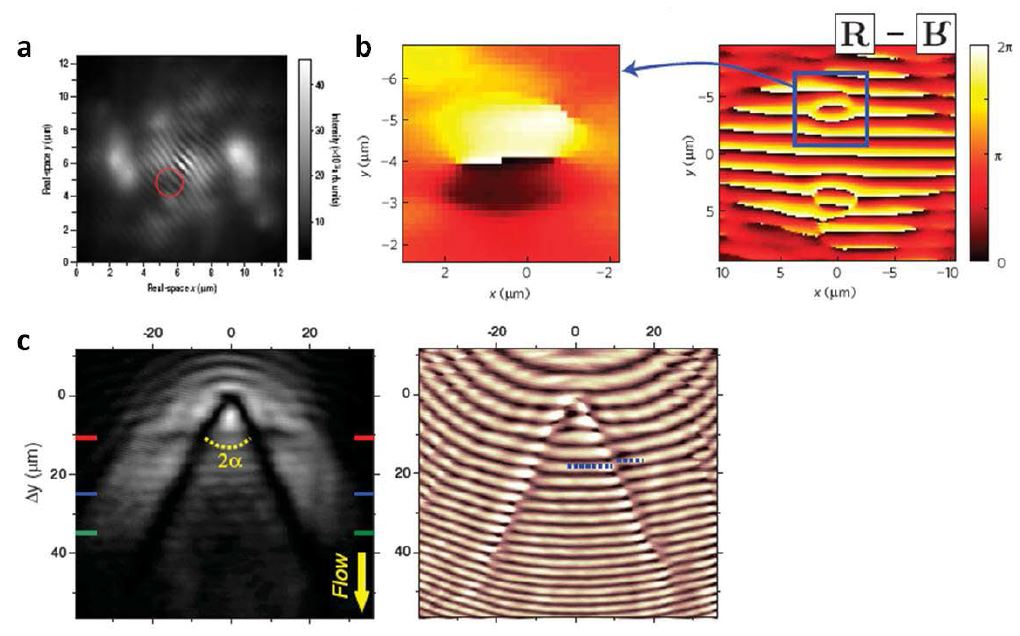}
\caption{\label{fig4} Vortices and solitons in exciton-polariton superfluids. {\bf a} Interferogram of a pinned vortex at a defect in a CdTe microcavity quantum well sample from Ref. \cite{lagoudakis08}. {\bf b} Experimental observation of a vortex-antivortex pair imposed by a laser pump spatial profile on a GaAs sample in Ref. \cite{roumpos11}. {\bf c} Intensity (left) and the phase (right) map of the propagating coherent polaritons across the barriers from Ref. \cite{amo11}. The solitons are seen as a phase slip.}
\end{figure}

\begin{figure}[t]
\includegraphics[width=\textwidth]{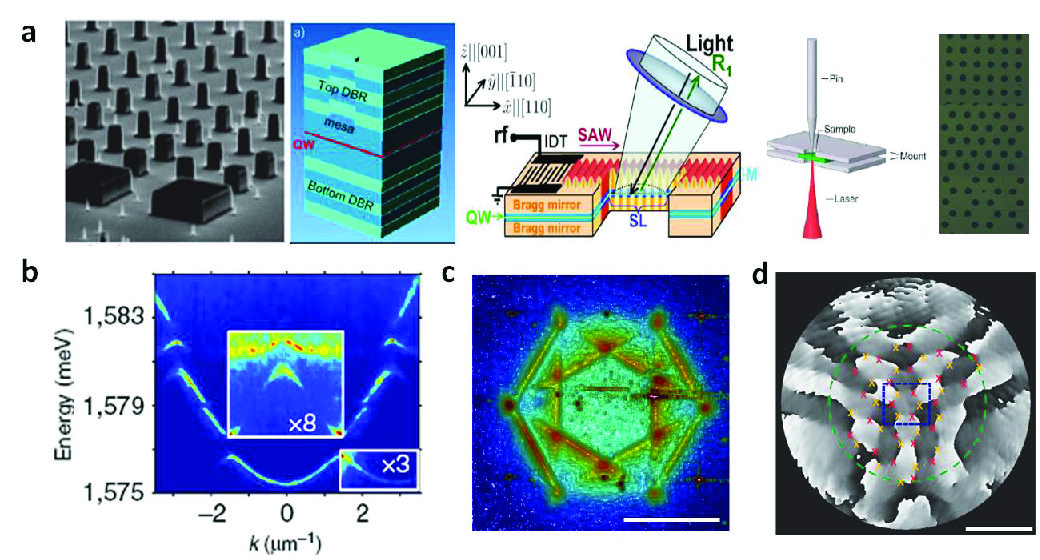}
\caption{\label{fig5}  Methods to create trapping potentials for exciton-polariton condensates. {\bf a} Various static and dynamical potential engineering methods: etching (first and second from left), surface acoustic wave (third), strain (fourth) and metal-film technique (rightmost). {\bf b} Experimentally observed one-dimensional band structure from a wire configuration in Ref. \cite{tanese13}. {\bf c} First three Brillouin zones measured for two-dimensional (2D) honeycomb lattices using the metal film technique. {\bf d} 2D lattice formation using laser spot engineering from Ref. \cite{tosi12}. }
\end{figure}

\end{widetext}

\end{document}